%% file: paper.tex
\documentclass{article}
\usepackage{spconf,amsmath,graphicx}
\include{macros}

\title{The role of asymptotic functions in network \\ optimization and feasibility studies}
\name{Renato L. G. Cavalcante and S.~Sta\'nczak\thanks{This work was partially supported by the Deutsche Forschungsgemeinschaft (DFG) under Grant STA 864/9-1 and by Nokia Bell Labs, Stuttgart, through the Nokia University Donation program.}}
\address{Fraunhofer Heinrich Hertz Institute and Technical University of Berlin\\
	email: \{renato.cavalcante, slawomir.stanczak\}@hhi.fraunhofer.de}
\begin{document}
\ninept
\maketitle
\begin{abstract}
Solutions to network optimization problems have greatly benefited from developments in nonlinear analysis, and, in particular, from developments in convex optimization. A key concept that has made convex and nonconvex analysis an important tool in science and engineering is the notion of asymptotic function, which is often hidden in many influential studies on nonlinear analysis and related fields. Therefore, we can also expect that asymptotic functions are deeply connected to many results in the wireless domain, even though they are rarely mentioned in the wireless literature. In this study, we show connections of this type. By doing so, we explain many properties of centralized and distributed solutions to wireless resource allocation problems within a unified framework, and we also generalize and unify existing approaches to feasibility analysis of network designs. In particular, we show sufficient and necessary conditions for  mappings widely used in wireless communication problems (more precisely, the class of standard interference mappings) to have a fixed point. Furthermore, we derive fundamental bounds on the utility and the energy efficiency that can be achieved by solving a large family of max-min utility optimization problems in wireless networks.
\end{abstract}
\section{Introduction}
\label{sec:intro}

Asymptotic functions have played a prominent role in the field of convex and nonconvex optimization \cite{aus03,rock70}, but many studies in optimization theory that have been influential to wireless communication engineers do not explicitly mention them, or these functions are used only in convex optimization problems. A fact that may partially explain the absence of asymptotic functions in developments in the wireless domain is that these functions are not necessarily easy to obtain, especially if we depart from the field of convex analysis. 

The first main contribution of this study is to show that the same analytic simplification used to compute asymptotic functions associated with convex functions can also be used to compute asymptotic functions associated with the so-called interference functions proposed in \cite{yates95}, which have found many applications in centralized and distributed solutions in wireless networks, including solutions to nonconvex problems \cite{martin11,slawomir09,renato2016,renato2016maxmin,renato2016power,feh2013,nuzman07,ho2015}. With this result, we construct simple self-mappings (hereafter called asymptotic self-mappings) associated with the mappings appearing in many of those studies.  

Asymptotic self-mappings set the stage for the second main contribution of the study. In more detail, we show that spectral properties of asymptotic self-mappings, which belong to a class widely investigated in the mathematical literature \cite{gau04,gun95,krause1986perron,krause01,nussbaum1986convexity}, are useful to obtain rigorous insights into properties of solutions to resource allocation and network feasibility problems within a unified framework. For example, we show that knowledge of the spectral radii of asymptotic mappings is useful to unify and generalize existing results in \cite{slawomir09,siomina12,renato2016,ho2014data} related to the feasibility of network topology designs (e.g., rank base stations according to the unserved traffic demand). In mathematical terms, we show a sufficient and necessary condition for the existence of fixed points of standard interference mappings that are not necessarily affine. 

As a second example of an application of asymptotic mappings, we build upon the results in \cite{nuzman07} to derive upper bounds for the utility and for the transmit energy efficiency (i.e., utility over power) achieved by solutions to utility maximization problems as a function of the power budget $\bar{p}$ available to transmitters. The bounds derived here are asymptotically tight and do not depend on any unknown constants, so they are particularly useful to determine whether a wireless network is likely to be noise limited or interference limited for a given power budget. They also provide us with guidance for the construction of good network topologies for large systems. In addition, they reveal that the network utility and the energy efficiency scale as $\Theta(1)$ and $\Theta(1/\bar{p})$, respectively, as $\bar{p}\to\infty$. We also obtain related technology-agnostic results for the case $\bar{p}\to 0^+$. All the theory developed here is verified in concrete network problems involving the well-known load coupled interference model investigated in, for example, \cite{renato2016maxmin,Majewski2010,renato14SPM,renato2016power,siomina2012load,ho2014data,ho2015,feh2013}. 

\section{Asymptotic mappings}
\label{sec:format}
Before deriving the first main contribution of this study, we briefly clarify notation and mathematical concepts. By $\real_+$ and $\real_{++}$ we denote the set of non-negative reals and positive reals, respectively.  Given $(\signal{x},\signal{y})\in\real^N\times\real^N$, vector inequalities such as $\signal{x}\le\signal{y}$ should be understood coordinate-wise. If $C\subset \real^N$ is a convex set, we say that a mapping $T:C\to\real^N:\signal{x}\mapsto[t_1(\signal{x}), \cdots, t_N(\signal{x})]$ is concave if the function $t_i:C\to\real$ is concave for every $i\in\{1,\ldots,N\}$. In this work, we use a slight modification of the original definition of standard interference functions in \cite{yates95}, which have found many applications in the wireless domain \cite{yates95,martin11,slawomir09,renato2016,renato2016maxmin,renato2016power,feh2013,nuzman07,ho2015}. In more detail, a function $f: \real^N \to \real_{++} \cup \{\infty\}$ is said to be a standard interference function if the following properties hold: ({\it scalability}) $(\forall \signal{x}\in\real^N_+)$ $(\forall \alpha>1)$  $\alpha {f}(\signal{x})>f(\alpha\signal{x})$, (b) ({\it monotonicity}) $(\forall \signal{x}_1\in\real_+^N)$ $(\forall \signal{x}_2\in\real_+^N)$ $\signal{x}_1\ge\signal{x}_2 \Rightarrow{f}(\signal{x}_1)\ge f(\signal{x}_2)$, and (c) $f(\signal{x})=\infty\Leftrightarrow \signal{x}\notin\real_{+}^N$. Given $N$ standard interference functions $t_i:\real^N \to \real_{++} \cup \{\infty\}$, $i=1,\ldots,N$, we call the mapping $T:\real^N_+\to\real_{++}^N:\signal{x}\mapsto[t_1(\signal{x}),\ldots, t_N(\signal{x})]$ a {\it standard interference mapping}. A norm $\|\cdot\|$ in $\real^N$ is said to be monotone if $(\forall\signal{x}\in\real^N)(\forall\signal{y}\in\real^N)$ $\signal{0}\le\signal{x}\le\signal{y}\Rightarrow\|\signal{x}\| \le \|\signal{y}\|$.

The fundamental mathematical tool used in this study is the analytic representation of asymptotic functions, which we state as a definition:

\begin{definition}
	\label{def.asymp_func} (\cite[Theorem~2.5.1]{aus03} Asymptotic function)
	The asymptotic function associated with a proper function $f:\real^N\to\real\cup \{\infty\}$ is the function given by
	\begin{align}
	\label{eq.asymp_func}
	\begin{array}{rcl}
	f_{\infty}:\real^N & \to & \real \cup \{\infty\} \\ 
	\signal{x} & \mapsto & \inf\left\{\displaystyle \liminf_{n\to\infty} \dfrac{f(h_n\signal{x}_n)}{h_n}~|~h_n\to\infty,~\signal{x}_n\to\signal{x} \right\},
	\end{array}
	\end{align} 
	where $(\signal{x}_n)_{n\in\Natural}$ and $(h_n)_{n\in\Natural}$ are sequences in $\real^N$ and $\real$, respectively. 
\end{definition}

Computing asymptotic functions directly from the definition is in general difficult, so many studies deal with only asymptotic functions associated with convex functions \cite{baus11,rock70}. The first main contribution of this study (Proposition~\ref{prop.af_properties}(i)) is to show that the same analytic simplification used for the computation of asymptotic functions $f_\infty$ associated with convex functions $f$ \cite[Corollary~2.5.3]{aus03} is also available if $f$ is a standard interference function that is not necessarily convex or concave. The proof of the next proposition is omitted owing to the space limitation, and we refer readers to the accompanying unpublished work in \cite{renato2017performance}.

\begin{proposition}
	\label{prop.af_properties}
	The asymptotic function $f_{\infty}:\real^N  \to  \real \cup \{\infty\}$ associated with a standard interference function $f:\real^N \to \real_{++} \cup \{\infty\}$ has the following properties:
	\begin{itemize}
		\item[(i)] $(\forall\signal{x}\in\real^N_+)~f_\infty(\signal{x}) = \lim_{h\to\infty} f(h \signal{x})/{h} \in\real_+$.
		
		\item[(ii)] $f_\infty$ is lower semicontinuous and positively homogeneous. If $f$ is in addition continuous when restricted to the non-negative orthant $\real_{+}^N$, then $f_\infty$ is continuous when restricted to the non-negative orthant $\real_{+}^N$.
		\item[(iii)] ({\it Monotonicity}) $(\forall \signal{x}_1\in\real_+^N)$ $(\forall \signal{x}_2\in\real_+^N)$ $\signal{x}_1\ge\signal{x}_2  \Rightarrow  {f}_\infty(\signal{x}_1)\ge f_\infty(\signal{x}_2)$.

	\end{itemize}
	
\end{proposition}

We now define the following mapping, which we later use to unify, generalize, and gain new insights into existing solutions in the wireless literature.

\begin{definition}
	\label{def.amap} (Asymptotic mappings)
	Let the function $t^{(i)}:\real^N\to\real_{++}\cup\{\infty\}$ be a standard interference function for each $i\in\{0,\ldots,N\}$. Given a standard interference mapping $T: \real^N_+ \to\real^N_{++}:\signal{x}\mapsto [t^{(1)}(\signal{x}),\cdots,t^{(N)}(\signal{x})]$, the asymptotic mapping associated with $T$ is given by 
	$
	T_\infty:\real^N_+\to\real^N_+:\linebreak[4]\signal{x}\mapsto [{t}^{(1)}_\infty(\signal{x}),\cdots,{t}^{(N)}_\infty(\signal{x})],
	$
	where, for each $i\in\{1,\cdots,N\}$, ${t}^{(i)}_\infty$ is the asymptotic function associated with $t^{(i)}$.
\end{definition}

 The properties proved in Proposition~\ref{prop.af_properties}(ii)-(iii) enable us to associate a common notion of spectral radius to continuous asymptotic mappings $T_\infty$. More precisely, following standard terminology in the mathematical literature \cite{nussbaum1986convexity}, we say that $\lambda\in\real_+$ and $\signal{x}\in\real^N_{+}\backslash\{\signal{0}\}$ are, respectively, an eigenvalue and an eigenvector of a continuous standard interference mapping or an asymptotic mapping $T$ if $T(\signal{x})=\lambda \signal{x}$. Furthermore, the spectral radius $\rho(T_\infty)$ of a continuous asymptotic mapping $T_\infty$ is the value given by $\rho(T_\infty) := \sup\{\lambda\in\real_+~|~(\exists \signal{x}\in\real_+^N\backslash\{\signal{0}\})~ T_\infty(\signal{x})=\lambda\signal{x}\} \in \real_+.$  Next, we show how these concepts naturally appear in the analysis of solutions to network optimization and network feasibility problems (not necessarily convex). We also show how to approximate the spectral radius with simple algorithms akin to the power method.

\section{Asymptotic mappings in wireless networks}
\subsection{Feasibility of network designs}

Let $\signal{X}\in\real_+^{N\times N}$ be a nonnegative matrix and $\signal{u}\in\real_{++}^N$ a positive vector. It is well known that the system $\signal{p}=\signal{u}+\signal{X}\signal{p}$ has a positive solution $\signal{p}=(\signal{I}-\signal{X})^{-1}\signal{u}\in\real_{++}^N$ if and only if the spectral radius of the matrix $\signal{X}$ is strictly less than one, which is a result that has played an important role in centralized and distributed resource allocation problems in wireless networks \cite[Ch.~2]{slawomir09}. Equivalently, we have $\signal{0}<\signal{p}=\signal{u}+\signal{X}\signal{p}\Leftrightarrow \signal{p}\in\mathrm{Fix}(T):=\{\signal{x}\in\real_{+}^N~|~T(\signal{x})=\signal{x}\}\ne\emptyset$, where $T$ is the standard interference mapping given by $T:\real^N_+\to\real^N_{++}:\signal{x}\mapsto \signal{X}\signal{x}+\signal{u}$. For this affine mapping $T$, note that its associated asymptotic mapping is $T_\infty:\real^N_+\to\real^N_{+}:\signal{x}\mapsto \signal{X}\signal{x}$, and thus the spectral radius of $T_\infty$ and the spectral radius of the matrix $\signal{X}$ (in the conventional sense in linear algebra) are the same. This observation suggests that, in feasibility analysis involving nonlinear mappings, the spectral radius of matrices should be replaced by the spectral radius of asymptotic mappings. We formally prove this fact in next proposition, which establishes the second main contribution of the study. The next proposition also shows a result that provides us with information about the location of the fixed point. The proof is also omitted owing to the space limitation.

\begin{proposition}
	\label{proposition.existence}
	Let $T:\real^N_+\to\real_{++}^N$ be a continuous standard interference mapping. Then $\mathrm{Fix}(T)\neq \emptyset$ if and only if $\rho(T_\infty)<1$. Furthermore, given an arbitrary monotone norm $\|\cdot\|$, there exists $\signal{x}^\star\in\mathrm{Fix}(T)\neq\emptyset$ with $\|\signal{x}^\star\|\le 1$ if and only if the tuple $(\signal{x}^\prime,\lambda^\prime)\in\real_{++}^N\times\real_{++}$ satisfying $T(\signal{x}^\prime)=\lambda^\prime\signal{x}^\prime$ and $\|\signal{x}^\prime\|=1$ is such that $\lambda^\prime \le 1$.~\footnote{The tuple $(\signal{x}^\prime,\lambda^\prime)$ exists and is unique, and it is easy to obtain with a simple fixed point algorithm \cite{nuzman07}.}
\end{proposition}
It is also worth mentioning that some distributed resource allocation algorithms (e.g., \cite{yates95,renato2016power}) have been designed under the assumption of the existence of the fixed point of a standard interference mapping. Proposition~\ref{proposition.existence} provides us with a complete characterization of its existence. Later, in Sect.~\ref{sect.examples}, we show that existing results related to the feasibility of OFDMA-based networks \cite{siomina12,ho2014data,renato2016} (among other applications) emerge as corollaries of  Proposition~\ref{proposition.existence}. In the next section we also show simple algorithms that are able to estimate the spectral radius of an arbitrary asymptotic mapping.

\subsection{Bounds on the solutions to utility optimization problems}
We now show that many properties of solutions to a large class of utility optimization problems can be explained by studying spectral properties of asymptotic mappings. In more detail, as shown in \cite{nuzman07,renato2016maxmin,sun2014,tan2014wireless} and the references therein, many (weighted max-min) utility maximization problems in wireless networks are particular instances of the following canonical optimization problem, originally shown in \cite{nuzman07} in the context of wireless networks: 
\begin{problem}
	\label{problem.canonical}
	(Canonical utility maximization problem)
	\begin{align}
	\label{eq.canonical}
	\begin{array}{lll}
	\mathrm{maximize}_{\signal{p}, c} & c \\
	\mathrm{subject~to} & \signal{p}\in \mathrm{Fix}(cT):=\left\{\signal{p}\in\real_{+}^N~|~\signal{p}=cT(\signal{p})\right\} \\
	& \|\signal{p}\|_a \le \bar{p} \\
	& \signal{p}\in\real_{+}^N, c\in\real_{++},
	\end{array}
	\end{align}
	where $\bar{p}\in \real_{++}$ is a design parameter hereafter called power budget,  $\|\cdot\|_a$ is an arbitrary monotone norm, and $T:\real_{+}^N\to\real_{++}^N$ is an arbitrary standard interference mapping. 
\end{problem}
Examples of problems that can be written in this canonical form include the max-min rate optimization in load coupled networks \cite{renato2016maxmin}, the joint optimization of the uplink power and the cell assignment \cite{sun2014}, the optimization of the uplink receive beamforming \cite[Sect.~1.4.2]{martin11}, and many of the applications described in \cite{cai2012optimal,tan2014wireless,zheng2016}. As shown in \cite{nuzman07}, Problem~\ref{problem.canonical} has a unique solution $(\signal{p}^\star,c^\star)\in\real_{++}^N\times \real_{++}$ that can be obtained with simple fixed point algorithms. In particular, $\signal{p}^\star$ is the limit of the sequence $(\signal{p}_n)_{n\in\Natural}\subset\real_{+}^N$ constructed according to $\signal{p}_{n+1}=(\bar{p} / \|{T}(\signal{p}_n)\|) {T}(\signal{p}_n)$, where $\signal{p}_1\in\real_{+}^N$ is arbitrary \cite{nuzman07}. Once $\signal{p}^\star$ is known, we can recover the optimal utility $c^\star$ by computing $c^\star = \bar{p}/\|{T}(\signal{p}^\star)\|$. Note that the fixed point iteration mentioned above is at the heart of some existing (semi)-distributed resource allocation algorithms (e.g., \cite{sun2015joint}), and we now prove results showing fundamental performance limits of the solutions obtained with these algorithms. 

 Since the solution to Problem~\ref{problem.canonical} exists for every $\bar{p}\in\real_{++}$, and it is unique, the following functions are well defined:

\begin{definition}
	\label{def.mm_ee}
	(Utility, power, and $\|\cdot\|_b$-energy efficiency functions) Denote by  $(\signal{p}_{\bar{p}},~c_{\bar{p}})\in\real_{++}^N\times\real_{++}$ the solution to Problem~\ref{problem.canonical} for a given power budget $\bar{p}\in\real_{++}$. The utility and power functions are defined by, respectively, $U:\real_{++}\to\real_{++}:\bar{p}\mapsto c_{\bar{p}}$ and $P:\real_{++}\to\real_{++}^N:\bar{p}\mapsto \signal{p}_{\bar{p}}$. In turn, given a monotone norm $\|\cdot\|_b$, the $\|\cdot\|_b$-energy efficiency function is defined by $E:\real_{++}\to\real_{++}:\bar{p}\mapsto U(\bar{p})/\|P(\bar{p})\|_b$. 
\end{definition}

By using the results in \cite{nuzman07}, we can show that the utility function $U$ and each coordinate of the power function $P$ are strictly increasing. We can also prove that the energy efficiency function is nonincreasing, and that all functions $U$, $P$, and $E$ are continuous in $\real_{++}$, but we omit the details owing to the space limitation. Our next main result shows that asymptotic functions naturally appear in the study of the behavior of these functions for sufficiently small and large values of the power budget. (See the accompanying unpublished work \cite{renato2017performance} for the proof of most results in Proposition~\ref{prop.acondv}.) 
\begin{proposition}
	\label{prop.acondv}
	Let $(\signal{p}_{\bar{p}}, 1/\lambda_{\bar{p}}):=(P(\bar{p}), U(\bar{p}))\in\real_{++}^N\times\real_{++}$ be the solution to Problem~\ref{problem.canonical} for a given power budget $\bar{p}\in\real_{++}$, and denote by $T_\infty:\real_{+}^N\to\real_{+}^N$ the asymptotic mapping associated with the standard interference mapping $T:\real_{+}^N\to\real_{++}^N$. Assume that $T$ is continuous and $\rho(T_\infty)>0$. Then,

	\item[(i)] $\lim_{\bar{p}\to\infty}\lambda_{\bar{p}}=\rho(T_\infty)=:\lambda_\infty$.
	\item[(ii)] Let the scalar $\lambda_\infty$ be as defined in (i), and denote by $(\bar{p}_n)_{n\in\Natural}\subset \real_{++}$ an arbitrary monotonically increasing sequence satisfying $\lim_{n\to\infty} \bar{p}_n = \infty$. Define $\signal{x}_{n} := (1/\|\signal{p}_{\bar{p}_n}\|_a)\signal{p}_{\bar{p}_n}$, and let $\signal{x}_{\infty}\in\real_+^N$ be an arbitrary accumulation point of the bounded sequence $(\signal{x}_n)_{n\in\Natural}\subset\real_{++}^N$. Then the tuple $(\signal{x}_\infty,\lambda_\infty)$ solves the following conditional eigenvalue problem:
	
	\begin{problem}
		\label{problem.eival_tinf}
		Find $(\signal{x}, \lambda)\in\real_{+}^N\times\real_{+}$ such that $T_\infty(\signal{x}) = \lambda \signal{x}$ and $\|\signal{x}\|_a = 1$.
	\end{problem}	
	\item[(iii)] $\sup_{\bar{p}>0} U(\bar{p}) = \lim_{\bar{p}\to \infty} U(\bar{p}) = 1/\lambda_\infty$ and $\sup_{\bar{p}>0} E(\bar{p}) = \lim_{\bar{p}\to 0^+} E(\bar{p}) = 1/\|T(\signal{0})\|_b$.
	\item[(iv)] $(\forall \bar{p}\in\real_{++})$ $ U(\bar{p})	\le  \min\{\bar{p}/\|T(\signal{0})\|_a, 1/\lambda_\infty\}$
	\item[(v)] $(\forall \bar{p}\in\real_{++}) E(\bar{p}) \le \min\{1/\|T(\signal{0})\|_b, \alpha/(\lambda_\infty~\bar{p})\}$, where $\alpha\in\real_{++}$ is any scalar satisfying $(\forall\signal{x}\in\real^N)~\|\signal{x}\|_a \le \alpha \|\signal{x}\|_b$ (such a scalar always exists because of the equivalence of norms in finite dimensional spaces).
	
	\item[(vi)] $U(\bar{p})\in\Theta(1)$ and $E(\bar{p})\in\Theta(1/\bar{p})$ as $\bar{p}\to\infty$.
	\item[(vii)] $U(\bar{p})\in\Theta(\bar{p})$ and $E(\bar{p})\in\Theta(1)$ as $\bar{p}\to 0^+$.
\end{proposition}

Before we explain in words the practical implications of the previous proposition, let us first define the following operating point, which has been motivated by the results in Proposition~\ref{prop.acondv}(iv):

\begin{definition}
	Assuming $\rho(T_\infty) = \lambda_\infty > 0$, we say that the network operates in the \emph{low power regime} if $\bar{p} \le \bar{p}_\mathrm{T}$ or in the \emph{high power regime} if $\bar{p} > \bar{p}_\mathrm{T}$, where the power budget $\bar{p}_\mathrm{T}:=\|T(\signal{0})\|_a/\lambda_\infty$ is called the \emph{transition point}.  
\end{definition}

In practice, as we will soon show in a numerical example, the transition point is the power budget in which networks are transitioning from a regime where the performance is \emph{limited by noise} to a regime where the performance is \emph{limited by interference.} For its determination, we only need to obtain the spectral radius of $T_\infty$. Under mild assumptions \cite{nuzman07,krause01,krause1986perron}, Problem~\ref{problem.eival_tinf} has a unique solution that can be solved with the same fixed point algorithm used to solve Problem~\ref{problem.canonical} (see its informal description above Definition~\ref{def.mm_ee}). However, even if Problem~\ref{problem.eival_tinf} does not necessarily have a unique solution, in which case the convergence of the fixed point algorithm described above may not be formally established with existing results, Proposition~\ref{prop.acondv}(i) provides us with a simple means to approximate the spectral radius of $T_\infty$ with any arbitrary precision. More precisely, we only need to solve Problem~\ref{problem.canonical} for $\bar{p}$ sufficiently large. By doing so, the reciprocal $1/c^\star$ of the optimal utility $c^\star$ is an approximation (more precisely, an upper bound) of the spectral radius of $T_\infty$. Additional details on the implications of Proposition~\ref{prop.acondv}(iii)-(vii) to resource allocation problems in wireless networks are presented in the next section, where we show a concrete application.

\section{Examples and final remarks}
\label{sect.examples}
We now apply the above results to feasibility and utility maximization problems based on the widely used load coupled interference model \cite{Majewski2010,renato14SPM,renato2016power,siomina2012load,ho2014data,ho2015,feh2013}. This model approximates the long-term behavior of modern communication systems (e.g., OFDMA-based systems), and it gives rise to mathematically tractable problems that have successfully addressed many system-level optimization tasks such as data offloading \cite{ho2014data}, load balancing and optimization \cite{siomina2012load,renato17load}, antenna tilt optimization \cite{feh2013}, energy savings \cite{ho2015,renato14SPM,renato2016power,pollakis12}, and rate optimization \cite{renato2016maxmin}, to cite a few. 

In the load coupling interference model, we divide the time and frequency grid into $K\in\Natural$ units called resource blocks. Users assigned to the same base station are not allowed to share resource blocks, but intercell interference is present because different base stations can allocate the same portion of the spectrum and time to serve their users. The set of $N$ users and $M$ base stations in the network is denoted by $\setn:=\{1,\ldots,N\}$ and $\setm:=\{1,\ldots,M\}$, respectively. The set $\setn_i\subset \setn$, assumed to be nonempty, is the set of users connected to base station $i\in\setm$. The pathloss between base station $i\in\setm$ and user $j\in\setn$ is given by $g_{i,j}\in\real_{++}$. The vector of transmit power and the load vector are given by, respectively, $\signal{p}=[p_1,\ldots,p_M]\in\real_{++}^M$ and $\signal{x}=[x_1,\ldots,x_M]\in\real_{++}^M$, where the $i$th coordinate of these vectors correspond to the power per resource block or the load at base station $i\in\setm$. Here, load is defined to be the fraction of resource blocks that a base station uses for data transmission. Note that this model assumes uniform transmit power per resource block, and it also assumes that all resource blocks experience the same (long-term) pathloss. If $\signal{p}\in\real^M_{++}$ is a fixed design parameter, the achievable rate of a resource block assigned by base station $i\in\setm$ to user $j\in\setn$ for a given load $\signal{x}\in\real^M_+$ is given by (see \cite{Majewski2010,renato14SPM,renato2016power,siomina2012load,ho2014data,ho2015,feh2013} for the limitations and strengths of the model):
\begin{align*}
\omega_{i,j}(\signal{x})=B\log_2\left(1+\dfrac{p_i g_{i,j}}{\sum_{k\in\setm\backslash\{i\}}{x}_k p_k g_{k,j}+\sigma^2}\right),
\end{align*}
where $\sigma^2\in\real_{++}$ is the noise per resource block and $B\in\real_{++}$ is the bandwidth of each resource block. By denoting by $d_j\in\real_{++}$ the data rate requested by user $j\in\setn$, the load at the base stations is obtained by computing the fixed point (if it exists) of the standard interference mapping given by   \cite{Majewski2010,siomina12,renato14SPM,feh2013}:
\begin{align}
\label{eq.load_mapping}
T:\real^M_{+}\to\real^M_{++}:\signal{x}\mapsto [ t^{(1)}(\signal{x}),\ldots,  t^{(M)}(\signal{x})]^T,
\end{align}
where, for each $i\in\setm$ and every $\signal{x}\in\real_+^M$, we define $t^{(i)}(\signal{x}) := \sum_{j\in\setn_i} {d_j}/({K\omega_{i,j}(\signal{x})}).$ We recall that $T$ has at most one fixed point \cite{yates95}. In the above formulation, components of the load vector can take values greater than one. In practice, the load at a base station cannot exceed one (otherwise the base station would be transmitting with more resources than available in the system), but knowledge of such values is useful to identify network bottlenecks by ranking base stations according to their unserved traffic demand \cite{siomina12}. Therefore, finding conditions for the existence of the fixed point of the mapping $T$ has been the focus of many studies \cite{Majewski2010,siomina12,renato2016,ho2015,renato14SPM}. Here we show that these results in the literature follow directly from Proposition~\ref{problem.eival_tinf}, and this proposition is also useful in the analysis of improved models for which existing results cannot be applied. More precisely, by using Proposition~\ref{prop.af_properties}(i), we can verify that the asymptotic mapping associated with $T$ is given by 
\begin{align}
\label{eq.load_mapping_asymptotic}
T_\infty:\real_{+}^M\to\real_{+}^M:\signal{x}\mapsto \mathrm{diag}(\signal{p})^{-1}\signal{M}\mathrm{diag}(\signal{p})\signal{x},
\end{align}
where $\mathrm{diag}(\signal{p})\in\real_+^{M\times M}$ is a diagonal matrix with diagonal elements given by the vector $\signal{p}$, and the component $[\signal{M}]_{i,k}$ of the $i$th row and $k$th column of the matrix $\signal{M}\in\real^{M\times M}_+$ is given by $[\signal{M}]_{i,k} = 0$ if $i= k$ or $[\signal{M}]_{i,k}= \sum_{j \in \mathcal{N}_i} {\mathrm{ln}(2)  d_j g_{k,j}}/({KB g_{i,j}})$ otherwise. By Proposition~\ref{proposition.existence}, we know that the load mapping $T$ has a fixed point if and only if $\rho(T_\infty)<1$. Since $T_\infty$ is the linear mapping shown in \refeq{eq.load_mapping_asymptotic}, we also know that the spectral radius $\rho(T_\infty)$ of $T_\infty$  and the spectral radius of the matrix $\mathrm{diag}(\signal{p})^{-1}\signal{M}\mathrm{diag}(\signal{p})$ (and hence of the matrix $\signal{M}$) coincide. Therefore, we conclude that $\mathrm{Fix}(T)\neq \emptyset$ if and only if the spectral radius $\rho(\signal{M})$ of the matrix $\signal{M}$ satisfies $\rho(\signal{M})<1$, which is exactly the result obtained in \cite{siomina12,ho2014data,renato2016} by using arguments with different levels of generality. This fact shows that all these existing results in the literature are unified and generalized by Proposition~\ref{proposition.existence}. None of these known results can be applied to more elaborate interference models where, for example, the rate of a user is upper bounded because of the limited number of modulation and coding schemes.  Such models have been considered in \cite{feh2013,renato14SPM}, and we now apply Proposition~\ref{proposition.existence} once again to study feasibility of a network design. More specifically, \cite{feh2013} (see also \cite{renato14SPM}) has proposed to replace the mapping $T$ in \refeq{eq.load_mapping} by the standard interference mapping $\bar{T}:\real^M_{+}\to\real^M_{++}:\signal{x}\mapsto [ \bar{t}^{(1)}(\signal{x}),\ldots,  \bar{t}^{(M)}(\signal{x})],$ where, for each $i\in\setm$,
\begin{align*}
\bar{t}^{(i)}:\real^M_+\to\real^M_{++}:\signal{x}\mapsto\sum_{j\in\setn_i} \max\left\{\dfrac{d_j}{K\omega_{i,j}(\signal{x})}, \dfrac{d_j}{u}\right\}
\end{align*}
and $u\in\real_+^M$ is the maximum rate that each resource block can achieve (because of the limited choice of modulation and coding schemes). The load is now the fixed point of the mapping $\bar{T}$, and its associated asymptotic mapping is also the mapping $T_\infty$ shown above; i.e.,
$\bar{T}_\infty:\real_{+}^M\to\real_{+}^M:\signal{x}\mapsto \mathrm{diag}(\signal{p})^{-1}\signal{M}\mathrm{diag}(\signal{p})\signal{x}$, where $\signal{M}$ is the matrix in \refeq{eq.load_mapping_asymptotic}. Therefore, by Proposition~\ref{proposition.existence}, we conclude that the improved mapping $\bar{T}$ has a fixed point if and only if $\rho(T_\infty)=\rho(\bar{T}_\infty)=\rho(\signal{M})<1$, which is the same criterion used to evaluate whether $T$ has a fixed point. This fact proves that existence of a feasible load vector does not depend on the choice of the modulation and coding schemes. 

\begin{figure}
	\begin{center}
		\includegraphics[width=0.8\columnwidth]{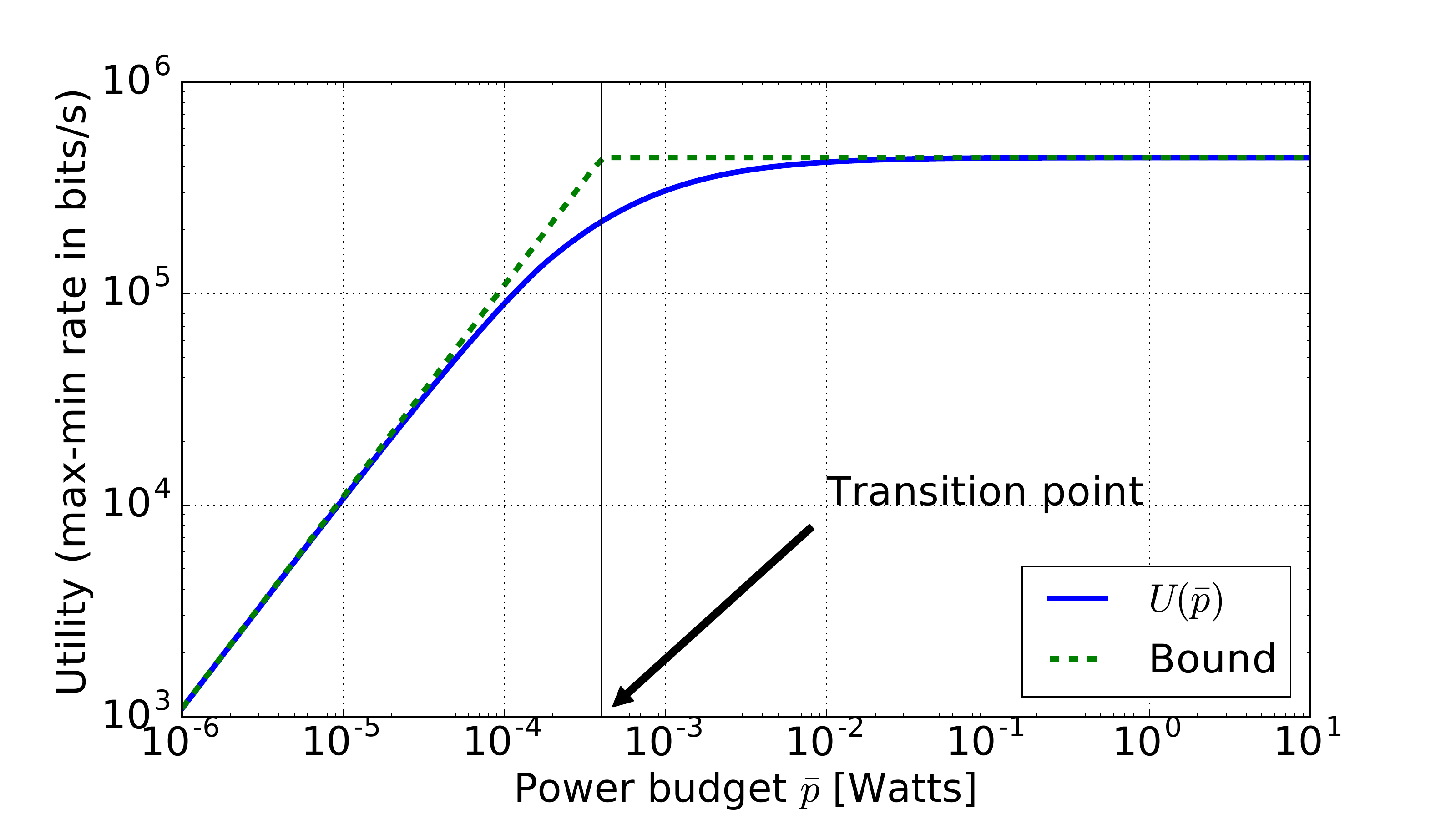}
		\caption{Network utility as a function of the power budget $\bar{p}$ for the problem described in \cite[Sect.~V-B]{renato2016maxmin}.}
		\label{fig.utility}
	\end{center}
	\vspace{-.6cm}
\end{figure}

\begin{figure}
	\begin{center}
		\includegraphics[width=0.8\columnwidth]{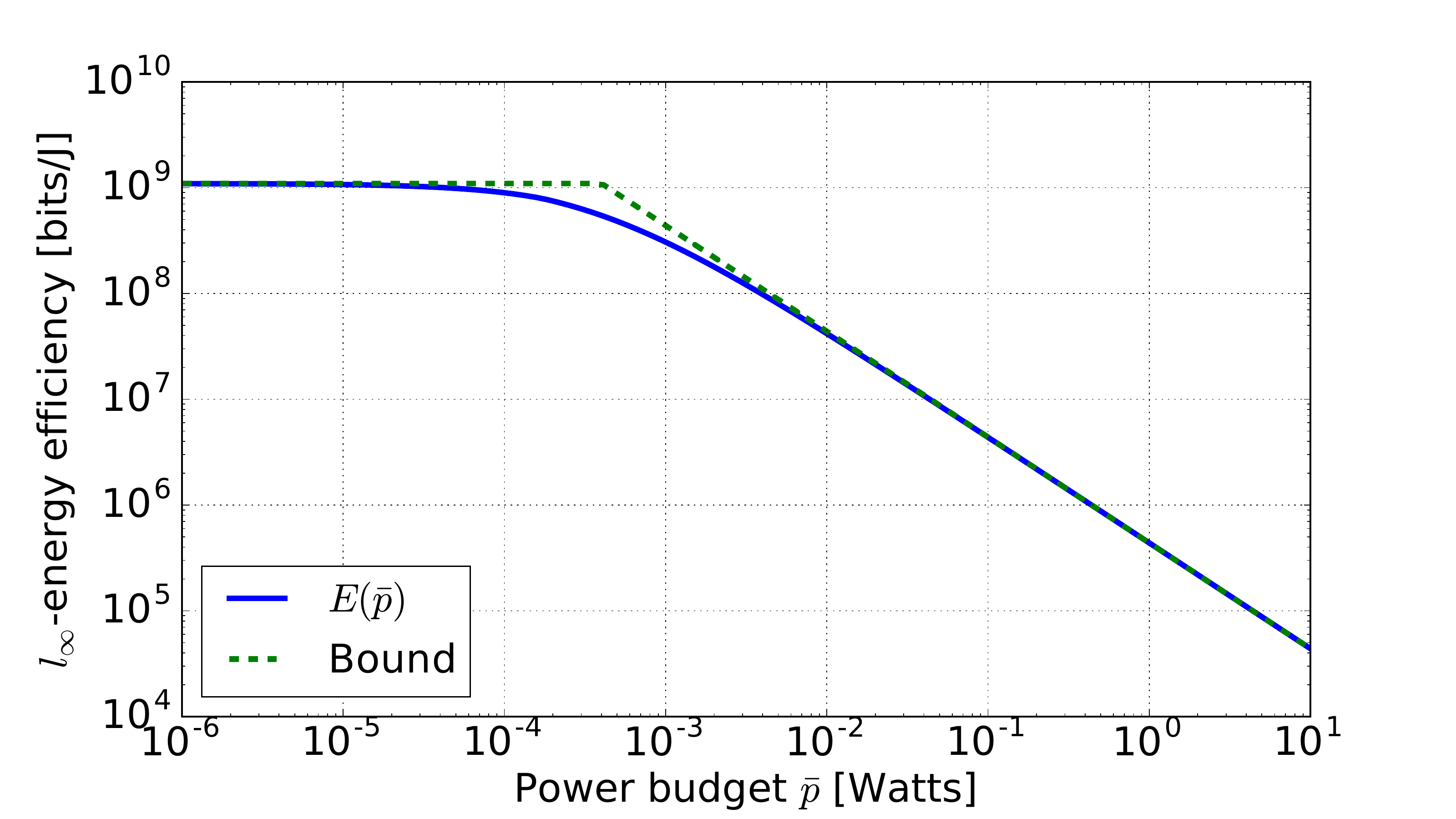}
		\caption{Energy efficiency as a function of the power budget $\bar{p}$ for the problem described in \cite[Sect.~V-B]{renato2016maxmin}.}
		\label{fig.energy_efficiency}
	\end{center}
\end{figure}

In the above analysis, both the users' rates and the power of base stations are given design parameters. We can also consider the problem of computing the transmit power and the corresponding load in order to maximize the minimum achievable rate of the users. In the context of the load coupled interference models described above, this utility maximization problem has been addressed in \cite{renato2016maxmin}, which has shown that, for optimality, all users should have the same rate, the load at all base stations should be set to one, and at least one base station should transmit with maximum power \cite[Proposition~2]{renato2016maxmin}.  Therefore, as shown in that study, the optimal rate and power allocation are solutions to a particular instance of Problem~\ref{problem.canonical}. Since details of the derivation can be found in \cite{renato2016maxmin}, we omit them for brevity. Here we show a numerical example illustrating the new insights gained with Proposition~\ref{prop.acondv}. Briefly, the numerical example shown in this study is similar to that used to produce \cite[Fig.~2]{renato2016maxmin}, with the only difference that here the noise power spectral density is fixed to -154~dBm/Hz, and we vary the power budget. Fig.~\ref{fig.utility} and Fig.~\ref{fig.energy_efficiency} show the utility (in bits/s) and the $\|\cdot\|_\infty$-energy efficiency (in bits/Joule) obtained in the simulations, respectively. All conditional eigenvalue problems have been solved with the fixed point iteration described just above Definition~\ref{def.mm_ee}. In the figures we verify that all technology-agnostic properties formally described by Proposition~\ref{prop.acondv} are present. For example, in the low power regime, the utility grows almost as a linear function, whereas the decay in energy efficiency is slow. These properties are a manifestation of Proposition~\ref{prop.acondv}(vii). In the high power regime, increasing the power budget by orders of magnitude brings only marginal gains in utility, an indication that the performance is limited by interference. Furthermore, the decay in energy efficiency is almost inversely proportional to the scaling of the power (e.g., doubling the power budget decreases the energy efficiency roughly by half). These results are a manifestation of the properties shown in  Proposition~\ref{prop.acondv}(vi). All bounds are asymptotically sharp as $\bar{p}\to\infty$ and as $\bar{p}\to 0^+$, which is a result consistent with Proposition~\ref{prop.acondv}(i)-(iii). We also verify that the transition point is a good indication of whether energy is being wasted. If we operate above the transition point, we can likely reduce the transmit power to obtain gains in energy efficiency at the cost of only a minor decrease in rates. In light of the results shown in this study, we should design networks in such a way that the transition point is larger than the power budget in general.

\vfill\pagebreak

\bibliographystyle{IEEEtran}
\bibliography{IEEEabrv,references}

\end{document}

%% file: macros.tex
\usepackage{ifpdf}
\usepackage{manfnt}
\usepackage[mathscr]{eucal}
\usepackage{graphicx}

\usepackage{algorithmicx}
\usepackage[ruled,vlined,commentsnumbered]{algorithm2e}
\usepackage{amssymb}
\usepackage{theorem}
\usepackage{caption}
\usepackage{subcaption}
\usepackage{cite}
\usepackage{color}
\usepackage{colortbl}
\definecolor{colorref}{rgb}{0.4648,0,0} 
\definecolor{colorcite}{rgb}{0,0.2902,0.1765}

\usepackage{bm}

\usepackage{amsmath}
\newtheorem{proposition}{Proposition}

\newcommand{\setn}{\mathcal{N}}
\newcommand{\setm}{\mathcal{M}}

\newcommand{\Rmnum}[1]{\uppercase\expandafter{\romannumeral #1}}

\newcommand{\signal}[1]{{\boldsymbol{#1}}}

\newcommand{\real}{{\mathbb R}}

\newtheorem{definition}{Definition}

\newtheorem{problem}{Problem}

\newcommand{\Natural}{{\mathbb N}}
\newcommand{\refeq}[1]{(\ref{#1})}

\newif\iftodo   
\todotrue
\newif\iftodoshort  
\todoshortfalse